\documentclass{aastex}

\def\farcs{\hbox{$.\!\!^{\prime\prime}$}}
\def\farc{\hbox{$\ \!\!^{\prime\prime}$}}

\slugcomment{To be submitted to the ApJL}

\shorttitle{Molecular Gas in 3C~31 and 3C~264}
\shortauthors{Lim et al.}

\begin{document}

\title{Molecular Gas in the Powerful Radio Galaxies 3C~31 and 3C~264: Major or Minor Mergers?}

\author{Jeremy Lim}
\affil{Academia Sinica Institute of Astronomy \& Astrophysics, Nankang, Taipei 11529, Taiwan}
\email{jlim@asiaa.sinica.edu.tw}

\author{Stephane Leon}
\affil{Academia Sinica Institute of Astronomy \& Astrophysics, Nankang, Taipei 11529, Taiwan}
\email{sleon@asiaa.sinica.edu.tw}

\author{Fran\b{c}oise Combes}
\affil{Observatoire de Paris, DEMIRM, 61 Avenue de l'Observatoire, 75014 Paris, France}
\email{Francoise.Combes@obspm.fr}

\and
\author{Dinh-V-Trung}
\affil{Academia Sinica Institute of Astronomy \& Astrophysics, Nankang, Taipei 11529, Taiwan}
\email{trung@asiaa.sinica.edu.tw}

\begin{abstract}
We report the detection of $^{12}$CO~($1 \rightarrow 0$) and $^{12}$CO~($2 \rightarrow 1$) emission from the central regions ($\lesssim 5$--$10 {\rm \ kpc}$) of the two powerful radio galaxies 3C~31 and 3C~264.  Their individual CO emission exhibits a double-horned line profile that is characteristic of an inclined rotating disk with a central depression at the rising part of its rotation curve.  The inferred disk or ring distributions of the molecular gas is consistent with the observed presence of dust disks or rings detected optically in the cores of both galaxies.  For a CO to H$_2$ conversion factor similar to that of our Galaxy, the corresponding total mass in molecular hydrogen gas is $(1.3 \pm 0.2) \times 10^9 {\rm \ M_{\odot}}$ in 3C~31 and $(0.31 \pm 0.06) \times 10^9 {\rm \ M_{\odot}}$ in 3C~264.  Despite their relatively large molecular-gas masses and other peculiarities, both 3C~31 and 3C~264, as well as many other powerful radio galaxies in the (revised) 3C catalog, are known to lie within the fundamental plane of normal elliptical galaxies.  We reason that if their gas originates from the mergers of two gas-rich disk galaxies, as has been invoked to explain the molecular gas in other radio galaxies, then both 3C~31 and 3C~264 must have merged a long time (a few billion years or more) ago but their remnant elliptical galaxies only recently (last tens of millions of years or less) become active in radio.  Instead, we argue that the cannibalism of gas-rich galaxies provides a simpler explanation for the origin of molecular gas in the elliptical hosts of radio galaxies.  Given the transient nature of their observed disturbances, these galaxies probably become active in radio soon after the accretion event when sufficient molecular gas agglomerates in their nuclei.  
\end{abstract}


\section{Introduction}
Bright radio sources served as the first signposts for highly energetic activity in galaxies.  The nature of these galaxies, and the reason for their luminous radio activity, have since been subjects of detailed investigation.  Although the vast majority resemble luminous elliptical galaxies, observations by  \citet{hec86} and \citet{smi89a,smi89b} showed that a significant fraction of the most powerful radio galaxies at low redshifts (henceforth meaning $z \lesssim 0.1$) exhibit peculiar optical morphologies suggestive of close encounters or mergers between galaxies.  Moreover, \citet*{smi90} showed that the visibly disturbed galaxies exhibit a slightly greater degree of rotational support than normal elliptical galaxies, and hence also are dynamically peculiar.  Nevertheless, all of the powerful radio galaxies examined by \citet{smi90}, mostly selected from the 3C catalog, lie within the fundamental plane of normal elliptical galaxies.  

Some radio galaxies possess so much dust that they can be detected in the far-infrared by {\it IRAS} \citep*{gol88}.  This discovery has motivated many subsequent searches for molecular gas in radio galaxies, in all cases by selecting those with known appreciable amounts of dust.  At low redshifts, eleven radio galaxies have so far been detected in $^{12}$CO \citep*{ric77,phi87,laz89,mir89,maz93,eva99,eva99a,eva99b}.  The detected galaxies span nearly three orders of magnitude in radio luminosity, and comprise those exhibiting core-dominated radio sources as well as classical double-lobed FR-I (edge darkened) and FR-II (edge brightened) radio sources.  All have inferred molecular-gas masses between $10^{9} {\rm \ M_{\odot}}$ and $10^{11} {\rm \ M_{\odot}}$, except for the very nearby radio galaxy Centaurus~A which has a molecular-gas mass of $\sim$$2 \times 10^{8} {\rm \ M_{\odot}}$ \citep{eck90}.  In four of the five cases (neglecting 3C~71, which is a spiral galaxy) mapped, Cen~A \citep{phi87,eck90}, B2~1318+343 \citep*{rad91}, 3C~293 \citep{eva99a}, and 4C~12.50 \citep{eva99b}, the CO gas is found to be concentrated in a compact (diameter of a few kpc or smaller) ring or disk around the center of the galaxy; in the fourth case, 3C~84 (Perseus~A), the gas is asymmetrically distributed over an extended (over 10~kpc) region, but also coinciding with the center of the galaxy \citep{ino96}.  The molecular gas in radio galaxies may therefore comprise the reservoir for fueling their central supermassive black holes.

At low redshifts, the only other type of galaxy to commonly exhibit molecular gas masses as high as $\ga 10^{10} {\rm \ M_{\odot}}$ are infrared-luminous galaxies.  These gas and dust rich galaxies exhibit vigorous star formation thought to be triggered by galaxy-galaxy interactions, and indeed the majority of ultraluminous infrared galaxies (i.e., those with $L(60{\mu}{\rm m}) \ga 10^{12} L_{\odot}$) are found to be merging systems of gas-rich disk galaxies.  In the latter systems, the CO gas is found to be preferentially concentrated in a ring or disk around the nuclear regions of the merging galaxies \citep{dow98,bry99}.  The observed similarities have prompted the suggestion that radio galaxies also originate from the mergers of two gas-rich disk galaxies, and in many cases comprise their still disturbed E/SO merger products \citep{mir89,maz93,eva99a,eva99b}.

Given the predisposition of the abovementioned surveys towards relatively dust-rich objects, are the radio galaxies so far detected biased towards those with unusually large amounts of molecular gas?  All the initial detections (except for 3C~71 and Centaurus~A) were made with the NRAO 12-m telescope, and in many cases reached lower limits just below the minimum gas mass detected.  As a result, it is not clear whether the IR-bright radio galaxies so far detected represent the most gas-rich members of a population that all possess substantial amounts of dust and gas, or extreme members of a population that possess a broad range of dust and gas masses.  This issue is of importance for a proper understanding of the nature of radio galaxies, and what fuels their supermassive black holes.

To address this issue, we have initiated a deep survey of all the previously undetected radio galaxies at redshifts $z \le 0.031$ in the revised 3C catalog \citep{spi85}.  The objects in this catalog represent the most luminous radio galaxies --- the majority of which exhibit classical double-lobed radio morphologies --- at their particular redshifts in the northern hemisphere; the selected objects therefore represent the most luminous radio galaxies in the Local Universe.  At low redshifts the vast majority can be clearly seen to be luminous elliptical galaxies, which in most cases comprise first ranked galaxies in poor clusters of roughly ten members although a number reside in much richer environments \citep[][and references therein]{zir96,zir97}.  Only a small minority have been detected in the infrared by {\it IRAS}; indeed, only three objects in this catalog have previously been detected in CO, one of which lies in the redshift range of our survey.

\section{Observations and Results}
We are conducting the survey in the $^{12}$CO ($1 \rightarrow 0$) and $^{12}$CO ($2 \rightarrow 1$) transitions simultaneously with the IRAM 30-m telescope, reaching lower limits in molecular-gas masses of typically $\sim$$1 \times 10^{8} {\rm \ M_{\odot}}$.  In the first phase of our survey from 1999~Dec~31 to 2000~Jan~3, when we observed nineteen radio galaxies, we detected 3C~31 and 3C~264, as well as a few other objects whose line profiles in one or both transitions need to be better measured.  Here, we present only the results for 3C~31 and 3C~264, both FR~I objects.  At the redshifted observing frequencies of $\sim$113~GHz and $\sim$226~GHz for both galaxies, typical system temperatures were $\sim$160~K for the $^{12}$CO ($1 \rightarrow 0$) transition and $\sim$340~K for the $^{12}$CO ($2 \rightarrow 1$) transition.  We used the filterbank spectrometer for the $^{12}$CO ($1 \rightarrow 0$) transition configured into two banks each having a channel bandwidth of 1~MHz spanning a total bandwidth of 512~MHz, and the autocorrelator for the $^{12}$CO ($2 \rightarrow 1$) transition configured into two sets each having a channel bandwidth of 1.25~MHz spanning a total bandwidth of 512~MHz.  For emission detected in the $^{12}$CO ($2 \rightarrow 1$) transition wider than the bandwidth spanned by the autocorrelator, as in the case of 3C~31, we subsequently configured the filterbank to span a total bandwidth of 1024~MHz for measuring this line.  In the observations, we used 50\farc\ throws on the wobbling secondary for sky measurements to measure the spectral baseline.

The line profiles of the $^{12}$CO ($1 \rightarrow 0$) and $^{12}$CO ($2 \rightarrow 1$) emissions detected from 3C~31 and 3C~264 are shown in Figure~1, along with the compact dust disks or rings observed in the cores of the respective radio galaxies with the HST by Martel et al. (1999).  Because the half-power beam of the IRAM 30-m telescope is 21\farcs2 /10\farcs6 at the $^{12}$CO ($1 \rightarrow 0$)/$^{12}$CO ($2 \rightarrow 1$) transitions observed here, we are mainly sensitive to emission from the central 7.8/3.4~kpc in 3C~31 ($z = 0.017$) and 9.8/4.9~kpc in 3C~264 ($z = 0.021$), where we have assumed $H_o = 65 {\rm \ km \ s^{-1} \ Mpc^{-1}}$ and ${\rm \Omega_o} = 1.0$ throughout this paper.  Note that the CO beam sizes are larger than the angular extents of the dust features seen in the respective galaxies.

\section{Properties of the Molecular Gas}
The double-horned line profile observed in each transition in each galaxy is characteristic of an inclined rotating disk with a central depression in CO emission at the rising part of its rotation curve \citep{wik97}.  The much wider (rotational) velocity width observed in 3C~31 ($\sim$$550 {\rm \ km \ s^{-1}}$) compared with 3C~264 ($\sim$$240 {\rm \ km \ s^{-1}}$) is consistent with the observed higher inclination of the dust disk or ring in 3C~31 ($\sim$$45^{\circ}$) compared with 3C~264 (nearly face on).  Corrected for inclination, the (maximum) rotational velocity of the molecular-gas disk/ring in 3C~31 is nearly $800 {\rm \ km \ s^{-1}}$, and is perhaps even higher in 3C~264; these rotation velocities are much higher than those seen in normal disk galaxies, let alone elliptical galaxies.  Such high rotation velocities, however, are commonly measured for the nuclear molecular-gas disks or rings in IR-luminous galaxies \citep{dow98,bry99}.

The integrated line intensities in both transitions for 3C~31 and 3C~264 are listed in Table~1.  The uncertainties quoted include an estimated uncertainty of $\sim$10\% in the absolute calibration of the temperature scale.  To convert from the integrated line intensity in $^{12}$CO ($1 \rightarrow 0$) to total mass in molecular hydrogen gas, we use the relationship $M(H_2) = 11.2 \times 10^{18} [(1+z-\sqrt{1+z})^2 / (1+z)] H_o^{-2} D^{-2} (K / \eta_A) \int T_a^* \ dv$ \citep*{gor92} as derived for molecular clouds in our Galaxy, where $D$ is the diameter of the IRAM 30-m telescope, $K$ a correction factor of order unity that depends on the size (and structure) of the source relative to the beam (here we assume $K=1$, as is appropriate when the source is much smaller than the beam), and $\eta_A = 0.57$ the forward efficiency of the telescope at our observing frequencies.  The derived molecular gas masses are $(1.3 \pm 0.2) \times 10^9 {\rm \ M_{\odot}}$ for 3C~31, a few times smaller than that in our Galaxy, and $(0.31 \pm 0.06) \times 10^9 {\rm \ M_{\odot}}$ for 3C~264, comparable only to that in Centaurus~A.


\section{Origin of Molecular Gas and Initiation of Radio Activity}
\subsection{Major Mergers}
By analogy with IR-luminous galaxies, mergers of two gas-rich disk galaxies provide an attractive explanation for the very large molecular-gas masses of $\ga 10^{10} {\rm \ M_{\odot}}$ and the compact circumnuclear distributions of this gas observed in a number of radio galaxies.  Direct evidence for such major mergers, as discussed by \citet{maz93}, can be seen in optical images of the two molecular-gas-rich radio galaxies B2~1318+343 and B2~1506+345; both systems appear to involve interactions between disk galaxies.  Furthermore, a possible evolutionary link between ultraluminous infrared galaxies and radio galaxies is provided by 4C~12.50 \citep{eva99}, which exhibits properties that place it in both categories; this galaxy shows two closely-separated nuclei that may indicate a late (major) merger.  If the remaining molecular-gas-rich radio galaxies also originate from major mergers, their singular nature suggest that they are observed at later stages.  Perhaps of significance then, the three radio galaxies to still show evidence for ongoing mergers all exhibit compact radio morphologies; many of the remaining galaxies, including 3C~31 and 3C~264, exhibit classical double-lobed radio morphologies (mostly FR~I objects).

By comparison with IR-luminous galaxies, however, radio galaxies exhibit a much broader range of molecular-gas masses.  Even among IR-bright radio galaxies, the larger fraction are not detectable with upper limits in molecular-gas masses as low as $\sim$$1 \times 10^{9} {\rm \ M_{\odot}}$ \citep{eva99,eva99b}.  By contrast, even relatively dim IR-luminous galaxies have molecular-gas masses $\ga 1 \times 10^{9} {\rm \ M_{\odot}}$ \citep*{san91}; moreover, as pointed out by \citet{shi98}, only the much brighter ultraluminous infrared galaxies, which typically have molecular-gas masses of $\sim$$10^{10} {\rm \ M_{\odot}}$, can possibly evolve into luminous elliptical galaxies.  Our (ongoing) survey imposes a more stringent upper limit on the molecular gas content of radio galaxies, suggesting that, even among the most powerful radio galaxies in the Local Universe, the majority are not detectable at molecular-gas masses of $\sim$$1 \times 10^{8} {\rm \ M_{\odot}}$.  To explain the relatively large range of molecular-gas masses seen just among the IR-bright radio galaxies, \citet{eva99b} suggest that "the consumption of molecular gas by extended star formation may be well underway prior to the onset of radio activity and may continue for another $10^7 {\rm \ yr}$ or so."  

The elevated star formation of IR-luminous galaxies is thought to last at most $\sim$$10^8 {\rm \ yr}$ \citep[e.g.,][]{mih94}, whereas the luminous radio lobes of radio galaxies are thought to have ages of just $\sim$$10^7$--$10^8 {\rm \ yr}$ \citep{blu00}, both very short by comparison with the typically $\sim$$10^9 {\rm \ yr}$ timescale of the merger process.  The suggestion that the luminous radio activity begins during or soon after the starburst phase, which itself begins before the galaxies completely merge, therefore poses a serious problem for radio galaxies such as 3C~31 and 3C~264, as well as 3C~84, 3C~120, 3C~293, which albeit rich in molecular gas and exhibiting other peculiarities lie within the fundamental plane of normal elliptical galaxies \citep{smi90}.  Studies show that merging systems of galaxies lie well away from the fundamental plane defined by elliptical galaxies \citep[e.g.,][]{jam99}, and that it takes a few billion years --- much longer than the merger timescale --- for the newly-formed stars to fade sufficiently for the merger remnant to evolve just in luminosity onto the fundamental plane \citep[e.g.,][]{shi98}.

Thus, if powerful radio galaxies such as 3C~31 and 3C~264 originate from disk-disk mergers, they must have formed at significantly higher redshifts than presently seen but only recently become active in radio.  For example, if it takes at least $\sim$$3 \times 10^9 {\rm \ yr}$ for the merging system to evolve into an elliptical galaxy \citep[e.g.,][]{shi98}, then in the adopted cosmology the progenitor galaxies of 3C~31 and 3C~264 must have merged at $z \ga 0.2$.  It is not obvious why the nuclei of such old merger remnants would suddenly become active long after much of the molecular gas has presumably been consumed or dispersed by star formation, unless this delay is related to how quickly the central black holes of the progenitor galaxies can coalesce to produce a rapidly-spinning supermassive black hole.  Such black holes my be necessary for powering luminous radio jets \citep[e.g.,][]{wil95}.

\subsection{Minor Mergers}
It is difficult to tell whether disturbed elliptical galaxies, particularly those like 3C~31 and 3C~264 betrayed largely or only by their compact gas and dust disks and having molecular-gas masses comparable with common disk galaxies, are the merger products of two gas-rich galaxies or have simply cannibalized smaller gas-rich galaxies.  In the latter scenario, the much more massive pre-existing elliptical galaxy can retain during the interaction its gross morphological and dynamical properties, yet exhibit disturbances tracing mostly the tidal rupture of the accreted galaxy.  This situation probably applies to the powerful radio galaxy Centaurus~A, which is believed to have accreted $\sim$$2 \times 10^{8} {\rm \ M_{\odot}}$ in molecular gas from a gas-rich galaxy $10^8$--$10^9 {\rm \ yr}$ ago \citep*{mal83,qui92}.  The observed disturbances in 3C~31 and 3C~264, both of which are larger and more luminous galaxies, resemble those seen in Centaurus~A.

In this picture, the majority of powerful radio galaxies at low redshifts are pre-existing (relatively massive) elliptical galaxies activated soon after an accretion event when sufficient gas agglomerates in their nucleus.  Their molecular gas content will then reflect the amount possessed by the accreted galaxy and the fraction that survives the merger process, and can therefore span a wide range of gas masses as is observed.  Because there are more lower-mass than higher-mass disk galaxies, one would expect such minor mergers to more often result in the accretion of relatively low molecular-gas masses.  In the cases of 3C~31 and 3C~264, the accretion of galaxies containing less molecular gas than our own Galaxy would suffice to explain their molecular gas content.

The potential greatest difficulty with this picture may be in explaining the very large molecular-gas masses of $\ga 10^{10} {\rm \ M_{\odot}}$ observed in radio galaxies such as 3C~84 (FR~I) and 3C~293 (FR~II) that lie within the fundamental plane of elliptical galaxies.  These relatively rare cases require the accretion of relatively massive gas-rich disk galaxies; although less common, such galaxies do exist.  Alternatively, multiple accretion events could be involved.

\acknowledgments
We are grateful to the staff of the IRAM 30-m telescope for a pleasant and productive observing session, particularly as our observations spanned the dawn of the popularly acclaimed new millenium and the awakening of the millenium bug.  J. Lim acknowledges the financial support of the National Science Council of Taiwan for conducting this research.

\clearpage

\begin{figure}
\vspace{-2cm}
\figurenum{1}
\epsscale{1.0}
\end{figure}
\vspace{-2cm}
\begin{figure}
\figurenum{1}
\epsscale{1.0}
\caption{\footnotesize
(See jpeg image) Top row: HST optical (negative) image of the center of 3C~31 (left panel), and the $^{12}$CO spectrum (at a velocity resolution of $20 {\rm \ km \ s^{-1}}$) measured towards the center of this galaxy as described in the text (right panel).  Bottom row: HST optical (negative) image of the center of 3C~264 (left panel), and the $^{12}$CO spectrum (velocity resolution of $15 {\rm \ km \ s^{-1}}$) measured towards the center of this galaxy (right panel).  The line in the upper right corner of the respective HST images denote the direction of the radio jet or lobes.  The $\pm 1\sigma$ error bar of each spectral profile is indicated in the respective panels.  Note that the ordinate for the $^{12}$CO ($2 \rightarrow 1$) transition has a range twice that of the corresponding $^{12}$CO ($1 \rightarrow 0$) transition.  The HST images are from \citet{mar99}.}
\end{figure}

\clearpage

\begin{deluxetable}{cccrrr}
\tablecaption{$^{12}$CO ($1 \rightarrow 0$) AND $^{12}$CO ($2 \rightarrow 1$) OBSERVATIONS OF 3C~31 AND 3C~264. \label{tbl-1}}
\tablewidth{0pt}
\tablehead{
\colhead{}                      & \colhead{}                            &
\colhead{}              &
\colhead{$\int T_a^* \ dv$}     &   
\colhead{$M$(H$_2$)}                \\
\colhead{NAME}                  & \colhead{REDSHIFT}                    &
\colhead{TRANSITION}            &
\colhead{$({\rm K \ km \ s^{-1}})$} & 
\colhead{$(M_{\odot})$}              
}
\startdata
3C~31   & 0.017 &  $^{12}$CO ($1 \rightarrow 0$) & $3.74 \pm 0.49$ &  $(1.3 \pm 0.2) \times 10^9$ \\
        &       &  $^{12}$CO ($2 \rightarrow 1$) & $8.44 \pm 1.14$  &                   \\
3C~264  & 0.022 &  $^{12}$CO ($1 \rightarrow 0$) & $0.55 \pm 0.10$ & $(0.31 \pm 0.06) \times 10^9$ \\
        &       &  $^{12}$CO ($2 \rightarrow 1$) & $0.84 \pm 0.17$ &                    \\
\enddata

\end{deluxetable}

\end{document}